\documentclass[10pt, aps, prl, twocolumn, superscriptaddress, showpacs]{revtex4-1}

\usepackage{here}
\usepackage{amsmath}
\usepackage{graphicx}
\usepackage{braket}
\usepackage{lettrine}
\usepackage{color}
\usepackage{times}
\usepackage{mathtools}

\begin{document}

\setlength{\textfloatsep}{15pt}
\setlength{\parskip}{0pt}

\title{
Intrinsic Spin and Orbital Hall Effects from Orbital Texture
}

\author{Dongwook Go}
\affiliation{Department of Physics, Pohang University of Science and Technology, Pohang 37673, Korea}

\author{Daegeun Jo}
\affiliation{Department of Physics, Pohang University of Science and Technology, Pohang 37673, Korea}

\author{Changyoung Kim}
\affiliation{Department of Physics and Astronomy, Seoul National University, Seoul 08826, Korea}

\author{Hyun-Woo Lee}
\email{hwl@postech.ac.kr}
\affiliation{Department of Physics, Pohang University of Science and Technology, Pohang 37673, Korea}



\begin{abstract}
We show theoretically that both intrinsic spin Hall effect (SHE) and orbital Hall effect (OHE) can arise in centrosymmetric systems through momentum-space orbital texture, which is ubiquitous even in centrosymmetric systems unlike spin texture. OHE occurs even without spin-orbit coupling (SOC) and is converted into SHE through SOC. The resulting spin Hall conductivity is large (comparable to that of Pt) but depends on the SOC strength in a nonmonotonic way. This mechanism is stable against orbital quenching. This work suggests a path for an ongoing search for materials with stronger SHE. It also calls for experimental efforts to probe orbital degrees of freedom in OHE and SHE. Possible ways for experimental detection are briefly discussed.
\end{abstract}

\maketitle


Spin Hall effect (SHE) \cite{Dyakonov1971, Hirsch1999, Engel2006, Schliemann2006, Niimi2015, Sinova2015} is a phenomenon in which an external electric field ${\bf E}$ generates a spin current in a transverse direction. 
When the spin current is injected to a neighboring ferromagnetic layer, it can even switch its magnetization direction~\cite{Liu2012_PRL,Liu2012_Science}.
SHE is now regarded as an indispensible tool in spintronics to generate and detect a spin current \cite{Niimi2015, Sinova2015}.
Of particular interest is  intrinsic mechanisms~\cite{Murakami2003, Sinova2004, Shen2004, Tanaka2008, Kontani2009}, which do not rely on impurity scattering. Large SHE in $5d$ transition metals such as Pt is believed to be intrinsic~\cite{Tanaka2008, Kontani2009, Guo2008, Freimuth2010, Morota2011, Sagasta2016,Yao2005}.


Spin-orbit coupling (SOC) is a crucial element for intrinsic SHE and has sizable value only near atomic nuclei \cite{Bihlmayer2006}, where it can be approximated as
\begin{equation}
H_{\rm so}=\frac{2\alpha_{\rm so}}{\hbar^2}{\bf S}\cdot {\bf L}.
\label{eq:H_SOC}{}
\end{equation}
Here, ${\bf L}$ denotes the orbital angular momentum near nuclei. Since the spin ${\bf S}$ couples to other degrees of freedom only through Eq.~(\ref{eq:H_SOC}) in non-magnets, ${\bf L}$ is expected to play important roles for SHE. Although an orbital degree of freedom is taken into account in equilibrium band structure calculations, {\it dynamical} roles of ${\bf L}$ are commonly ignored in literature.
Only for a limited class of systems, it was argued~\cite{Kontani2008,Tanaka2008,Kontani2009,Tanaka2010}  that an Aharonov-Bohm phase generated by orbitals is responsible for SHE and that SHE is closely related to orbital Hall effect (OHE). Here, OHE refers to an ${\bf E}$-induced transverse flow of ${\bf L}$ \cite{orbital_current_definition}. 
Even for these materials, however, there is no experiment that probes roles of ${\bf L}$ as far as we are aware of. It is partly due to the expectation that ${\bf L}$ cannot play any important roles due to orbital quenching~\cite{Kittel_textbook} in solids.


For centrosymmetric systems with  momentum-space orbital texture, we demonstrate that  ${\bf E}$ generates nonzero ${\bf L}$ (even when ${\bf L}$ is quenched in equilibrium), which leads to intrinsic SHE and OHE since the generated ${\bf L}$ is odd in the crystal momentum ${\bf k}$. This mechanism provides not only an alternative theoretical picture to understand SHE and OHE in Refs.~\cite{Kontani2008,Tanaka2008,Kontani2009} but also implies that many other systems may exhibit SHE and OHE since  the orbital texture is ubiquitous in multi-orbital systems.
Specifically we demonstrate two points: (i) even when SOC is absent and ${\bf L}$ is completely  quenched in equilibrium, the orbital texture generates OHE universally. (ii) When  SOC is sizable,  OHE is efficiently converted into SHE. Thus OHE is more fundamental than SHE in this mechanism.
Interestingly we find that stronger SOC does not necessarily imply enhanced SHE. This result is relevant for ongoing search for materials with strong SHE.

\begin{figure}[t!]
\includegraphics[angle=0, width=0.47\textwidth]{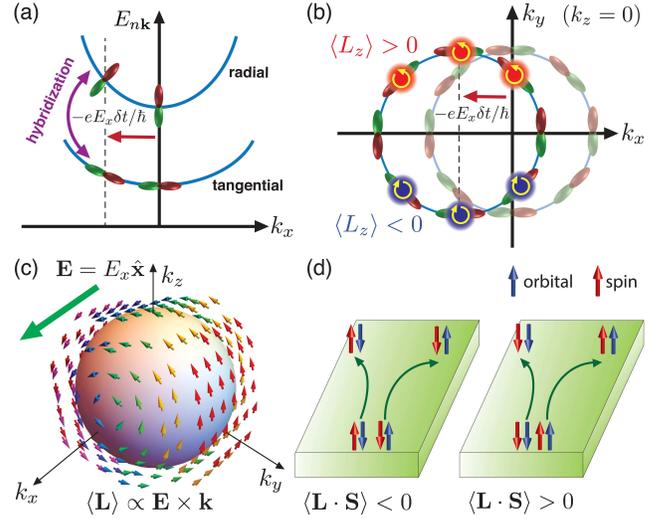}
\caption{(a,b) Illustration of intrinsic OHE from orbital texture in centrosymmetric systems. SOC is ignored for simplicity. (a) Schematic band structure with plots of wavefunction character at each band. Here, $k_y=k_z=0$. (b) When an electron in the lower band is pushed from ${\bf k}$ to ${\bf k}+\delta{\bf k}$ by an external electric field ${\bf E}\parallel +\hat{\mathbf{x}}$, positive(negative) $\langle {L_z} \rangle$ is induced for the non-equilibrium state with $k_y>0$ ($k_y<0$). (c) In three-dimensional $\mathbf{k}$-space, $\left\langle \mathbf{L} \right\rangle$ is induced into the direction of $\mathbf{E}\times\mathbf{k}$. This leads to finite $\left\langle L_z v_y \right\rangle$, OHE.
(d) When SOC is taken into account,  SHE occurs in the same or opposite direction of OHE depending on whether the correlation $\left\langle {\bf L}\cdot {\bf S} \right\rangle$ is positive or negative.
}
\label{fig:illustration}
\end{figure}

We begin with an illustration of the point (i) for a $p$-orbital system. We assume $\alpha_{\rm so}=0$ since  SOC is not essential for (i). We also assume that all orbital degeneracy is lifted and the expectation value of ${\bf L}$ is suppressed to zero for all eigenstates. Nevertheless, the orbital texture can be present; the orbital character may vary with ${\bf k}$ and from bands to bands.
For concreteness, we assume that for ${\bf k}=|{\bf k}|(\cos\phi,\sin\phi,0)$ in the $k_z=0$ plane, the wavefunction in the upper(lower) band has radial(tangential) $p$-orbital character, that is $|u^\textup{upper}_{{\bf k}}\rangle\sim|p_{\phi}\rangle$
($|u^\textup{lower}_{{\bf k}}\rangle\sim|p_{\phi+\pi/2}\rangle$) [Fig. 1(a)].
Here
$|u^\textup{upper(lower)}_{{\bf k}}\rangle$ is the periodic part of the Bloch wavefunction of the upper(lower) band and $|p_\phi\rangle\equiv \cos \phi |p_x\rangle + \sin \phi |p_y\rangle$.
Figure~1(b) shows schematically the wavefunction character of the eigenstates in the lower band at the Fermi surface.
Note that the expectation value $\langle {\bf L} \rangle$ vanishes for each of these states. Suppose ${\bf E}=E_x\hat{\bf x}\ (E_x>0)$ is then applied to shift \color{black} $\mathbf{k}$ to ${\bf k}+\delta{\bf k}=|{\bf k}+\delta{\bf k}|(\cos(\phi+\delta \phi),\sin(\phi+\delta \phi),0)$, where $\delta\phi$ is positive(negative) for positive(negative) $k_y$.
Under this ${\bf k}$ shift, $|u^\textup{lower}_{{\bf k}}\rangle\sim|p_{\phi+\pi/2}\rangle$, which can be written as $|p_{\phi+\pi/2}\rangle=|p_{\phi+\delta\phi+\pi/2}\rangle+\delta\phi|p_{\phi+\delta\phi}\rangle$, evolves with time to $\exp(-iE^\textup{lower}_{{\bf k}+\delta{\bf k}}\delta t/\hbar)|u^\textup{lower}_{{\bf k}+\delta{\bf k}}\rangle +\delta\phi \exp(-iE^\textup{upper}_{{\bf k}+\delta{\bf k}}\delta t/\hbar)|u^\textup{upper}_{{\bf k}+\delta{\bf k}}\rangle$. Thus,  an interband superposition \color{black} is induced by ${\bf E}$.
Note that the ratio between the two coefficients of the states $|u^\textup{upper/lower}_{{\bf k}+\delta{\bf k}}\rangle$ is {\it complex}, implying that the superposition contains the component $|p_x\rangle \pm i |p_y\rangle = |L_z=\pm \hbar \rangle$ and its expectation value $\langle L_z \rangle$ is {\it nonzero}. 
Thus, even when ${\bf L}$ is completely quenched in equilibrium, dynamically induced interband superpositions can have nonzero $\langle{\bf L}\rangle$.
An explicit calculation results in $\langle {\bf L} \rangle \propto \delta\phi\hat{\bf z}$,
which points in opposite directions for positive and negative $k_y$'s [Fig.~\ref{fig:illustration}(b)].
This two-dimensional profile of $\langle {\bf L}\rangle$ in the $k_z=0$ plane can be easily extended to a three-dimensional one by rotating Fig.~\ref{fig:illustration}(b) around ${\bf E}$. Figure~\ref{fig:illustration}(c) shows schematically the resulting three-dimensional profile of $\langle {\bf L}\rangle \propto {\bf E}\times {\bf k}$.
Note that although the sum of $\langle {\bf L} \rangle$ over occupied superposed states may vanish, the sum of the orbital Hall current $\sim \langle v_y L_z \rangle$ is nonzero.
This illustrates an intrinsic mechanism of OHE based on the orbital texture.
By the way, the $\langle {\bf L} \rangle$ profile in Fig.~\ref{fig:illustration}(c) is similar to the {\it equilibrium} $\langle {\bf L} \rangle$ profile in orbital Rashba systems~\cite{Park2011,Park2015}
despite the crucial difference that $\langle {\bf L} \rangle$ in Fig.~\ref{fig:illustration}(c) is evaluated for dynamically induced {\it nonequilibrium} interband superpositions whereas $\langle {\bf L} \rangle$ in orbital Rashba systems for {\it equilibrium} eigenstates.
\color{black}

Next we restore SOC. Then due to the correlation between ${\bf L}$ and ${\bf S}$, nonzero $\langle \mathbf{L} \rangle$ in Fig.~\ref{fig:illustration}(c) implies $\langle \mathbf{S} \rangle$ to be nonzero.
Thus SOC generates SHE as a {\it concomitant} effect of OHE. The sign of the spin Hall conductivity (SHC) is the same or opposite to that of the orbital Hall conductivity (OHC) depending on whether the correlation is positive or negative (that is, ${\bf S}$ is parallel or antiparallel to ${\bf L}$) [Fig.~1(d)].

\begin{figure}[t!]
\includegraphics[angle=0, width=0.49\textwidth]{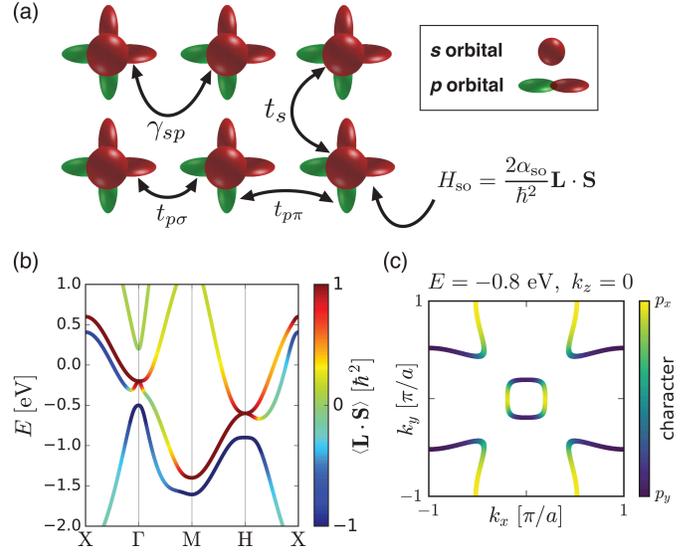}
\caption{
(a) A tight-binding model on a simple cubic lattice with $s$, $p_x$, $p_y$, and $p_z$ orbitals at each site. The nearest-neighbor hopping amplitude between $s$ orbitals is $t_s$, and that between $p$ orbitals via $\sigma(\pi)$ bonding is $t_{p\sigma(\pi)}$. An inter-orbital hopping amplitude from $p_x$, $p_y$, or $p_z$ orbital to $s$ orbital is $\gamma_{sp}$.
(b) Band structure obtained from the tight-binding model. The color represents the correlation $\left\langle \mathbf{L}\cdot\mathbf{S} \right\rangle$ for each eigenstate.
(c) Wavefunction character of the eigenstates at $E=-0.8$ eV with $k_z=0$.
}
\label{fig:model}
\end{figure}

The orbital texture assumed in Fig.~\ref{fig:illustration}(a) occurs even in very trivial systems. To demonstrate this point, we adopt a tight-binding model description of a simple cubic lattice with four orbitals $s$, $p_x$, $p_y$, $p_z$ at each lattice point. Possible nearest-neighbor hoppings are shown in Fig.~2(a) with their hopping amplitudes (see Ref.~\cite{Supplementary} for details).
Figure~2(b) shows the band structure of this system. The three doubly-degenerate lower bands have $p$-character whereas the topmost doubly-degenerate band (with $\Gamma$ point band edge at $0.3$ eV) has $s$-character.
Figure~2(c) shows the orbital character of the states at $E=-0.8$ eV in the $k_z=0$ plane.
Note that the inner(outer) band has radial(tangential) character orbital texture as assumed in Fig.~\ref{fig:illustration}(a).
We emphasize that the orbital texture arises from the $sp$ hopping $\gamma_{sp}$, which mediates the ${\bf k}$-dependent hybridization between $p_x$, $p_y$, and $p_z$ orbitals in  eigenstates. When $\gamma_{sp}=0$,  the orbital texture disappears. Thus $\gamma_{sp}$ may be regarded as a measure of the orbital texture strength.
Numerical values of the Hamiltonian parameters are (unless mentioned otherwise) as follows; $E_s = 3.2$, $E_{p_x}=E_{p_y}=E_{p_z}=-0.5$ for on-site energies, $t_s = 0.5$, $t_{p\sigma} = 0.5$, $t_{p\pi} = 0.2$, $\gamma_{sp} = 0.5$ for nearest-neighbor hopping amplitudes, and $\alpha_\textup{so} = 0.1$ for SOC, all in unit of $\mathrm{eV}$.

To assess the role of the orbital texture for OHE and SHE rigorously, one should go beyond the crude evaluation of the interband superposition given above, which captures only the {\it initial} evolution of an eigenstate toward its nonequilibrium steady state. For this, we use the Kubo formula,
\begin{subequations}
\label{eq:Kubo_1}
\begin{align}
\label{eq:Kubo_1a}
&\sigma_\textup{OH(SH)}
=
\frac{e}{\hbar}  \sum_{n\ne m} \int \frac{d^3k}{(2\pi)^3}  \left(  f_{m\mathbf{k}}-f_{n\mathbf{k}}  \right) \Omega_{nm\mathbf{k}}^{X_z},
\\
\label{eq:Kubo_1b}
&\Omega_{nm\mathbf{k}}^{X_z}
=
\hbar^2
\textup{Im}
\left[
\frac{
\bra{u_{n\mathbf{k}}}
 j_{y}^{X_z}
\ket{u_{m\mathbf{k}}}
\bra{u_{m\mathbf{k}}}
 v_x
\ket{u_{n\mathbf{k}}}
}
{ \left( E_{n\mathbf{k}} - E_{m\mathbf{k}} + i\eta \right)^2}
\right],
\end{align}
\end{subequations}
to calculate OHC ($\sigma_{\rm OH}$) and SHC ($\sigma_{\rm SH}$) for the tight-binding model with the orbital texture. Here \color{black}
$ j_y^{X_z}  = \left(  v_y  X_z  + X_z  v_y   \right)/2$ is the conventional orbital(spin) current operator with $X_z=L_z(S_z)$, $v_{x(y)}$ is the velocity operator along the $x(y)$ direction, 
and $f_{n\mathbf{k}}$ is the Fermi-Dirac distribution function.
In view of the illustration in Fig.~1, $\Omega_{nm\mathbf{k}}^{X_z}$ amounts to the contribution to $\sigma_{\rm OH(SH)}$ from the interband superposition between the bands $n$ and $m$. Figures~3(a) and 3(b) show respectively the calculated $\sigma_\mathrm{OH}$ and $\sigma_\mathrm{SH}$ as a function of the Fermi energy $E_{\rm F}$ for different orbital texture strengths.
Note that both $\sigma_\mathrm{OH}$ and $\sigma_\mathrm{SH}$ vanish for $\gamma_{sp}=0$ and grow with increasing $\gamma_{sp}$. Thus the orbital texture is crucial not only for OHE but also for SHE. Note that for $\gamma_{sp}\gtrsim$ 0.1 eV, both $\sigma_{\rm OH}$ and $\sigma_{\rm SH}$ can be gigantic $\sim 10^3$ $\hbar/2|e|$ $(\Omega \cdot {\rm cm})^{-1}$,
which is comparable to SHC of Pt \cite{Kimura2007, Guo2008, Sagasta2016}.

\begin{figure}[t!]
	\includegraphics[angle=0, width=0.43\textwidth]{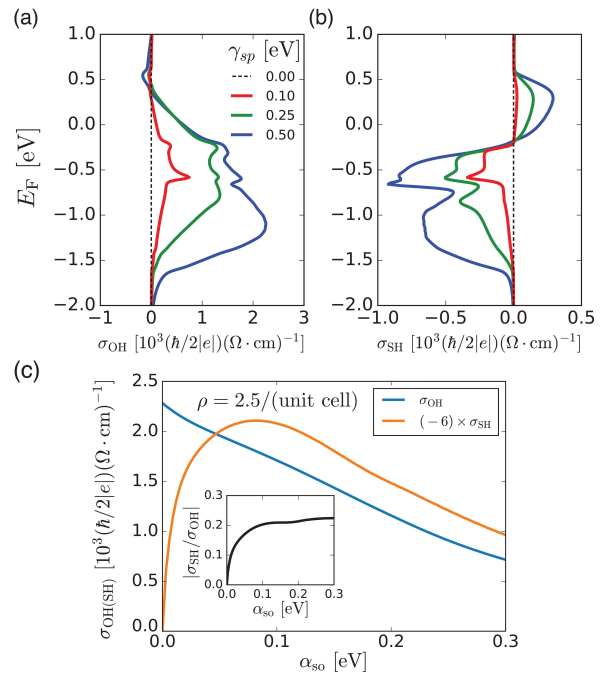}
	\caption{
		The $E_{\rm F}$ dependences  of  (a) OHC $\sigma_\mathrm{OH}$  and  (b) SHC $\sigma_\mathrm{SH}$ for different values of the $sp$ hopping amplitude $\gamma_{sp}$. (c) The SOC dependences of $\sigma_\mathrm{OH}$ and $\sigma_\mathrm{SH}$ for a fixed electron density, $\rho=2.5/\textup{(unit cell)}$. Inset: Conversion efficiency $\sigma_\mathrm{SH}/\sigma_\mathrm{OH}$ as a function of $\alpha_\mathrm{so}$.
	}
\label{fig:Hall-conductivities}
\end{figure}

Figure. 3(c) shows the SOC strength ($\alpha_{\rm so}$) dependence of $\sigma_\mathrm{OH}$ and $\sigma_\mathrm{SH}$
(see Ref.~\cite{Supplementary} for further details) for a fixed electron density 
$\rho=2.5$ electrons per unit cell, which corresponds to
$E_{\rm F}\approx -0.7$ eV though precise $E_{\rm F}$ value varies with $\alpha_{\rm so}$. 
Note that $\sigma_\mathrm{OH}$ is nonzero even when $\alpha_{\rm so}=0$, confirming that OHE can arise even without SOC. On the other hand, $\sigma_\mathrm{SH}=0$ for $\alpha_{\rm so}=0$ and increases as $\alpha_{\rm so}$ is turned on. These results are consistent with the interpretation that OHE arises first and SHE is converted from OHE through SOC.
An additional support to this interpretation comes from microscopic ($\mathbf{k}$- and band-resolved) contributions to $\sigma_{\rm OH}$ and $\sigma_{\rm SH}$~\cite{Supplementary},
which \color{black} qualitatively match with each other once the correlation $\left \langle {\bf L}\cdot{\bf S} \right\rangle$ distribution in Fig.~2(b) is taken into account.
Interestingly $\sigma_{\rm SH}$ decreases when $\alpha_{\rm so}$ goes beyond a threshold value ($\sim 0.1$ eV).
Such nonmonotonic dependence on $\alpha_{\rm so}$ can be understood as a combined effect of two trends:
enhanced SOC may reduce $\sigma_{\rm OH}$ [Fig.~\ref{fig:Hall-conductivities}(c)]
and the conversion efficiency from OHE to SHE, $\sigma_{\rm SH}/\sigma_{\rm OH}$ [inset in Fig.~\ref{fig:Hall-conductivities}(c)], saturates once a system enters the strong SOC regime.
This result implies that materials with overly strong SOC may not be good choices for large SHC. 

Interestingly, OHC and SHC remain stable against crystal field splitting.
When the on-site energies of $p_x$ and $p_y$ orbitals are shifted by $\pm\Delta_{\rm cf}$, respectively,
we find~\cite{Supplementary} that
$\sigma_{\rm OH(SH)}$ remains intact even for $\Delta_{\rm cf}$ as large as 0.3 eV because orbital degeneracy between $p$-character bands is already lifted by $\gamma_{sp}$
for most ${\bf k}$ points except a few high symmetry points, such as $\Gamma$ and $\mathrm{H}$ [Fig.~\ref{fig:model}(b)].
\color{black}

%

Next we compare our work with other theoretical works. For two-dimensional Rashba systems, Ref.~\cite{Sinova2004} reports that momentum-space spin texture generates an interband superposition upon the application of ${\bf E}=E_x\hat{\bf x}$ and $\langle S_z\rangle$ for the superposition has opposite signs for opposite signs of $k_y$'s. This mechanism (Fig.~1 of Ref.~\cite{Sinova2004}) is thus very similar to ours (Fig.~1 of this Letter) and our work may be regarded as an orbital counterpart of Ref.~\cite{Sinova2004}. There are clear differences, however. Reference~\cite{Sinova2004} completely ignores the orbital degree of freedom and is applicable only to noncentrosymmetric systems, whereas ours is applicable to centrosymmetric systems. The result in Ref.~\cite{Sinova2004} is significantly affected by the vertex correction~\cite{Inoue2004}, which captures effects of impurity scattering, whereas
our result is not
since the vertex corrections for $\sigma_{\rm SH}$ and $\sigma_{\rm OH}$ vanish in the weak scattering limit due to the inversion symmetry~\cite{Bernevig2005, Tanaka2008,Murakami2004PRBb}.

For $4d$ and $5d$ transition metals, Refs.~\cite{Tanaka2008,Kontani2009} report 
the inter-atomic hopping between $s$ and $d$ orbitals 
to be important for OHE and SHE and interpret the result in terms of the Aharonov-Bohm phase. We argue that the result may be interpreted alternatively as a $d$ orbital version of the orbital-texture-based mechanism (Fig.~\ref{fig:illustration}). In centrosymmetric transition metals of fcc or bcc crystal structure, the $sd$ hopping between nearest neighbor atoms generates orbital texture in $d$-character bands. Then for ${\bf E}=E_x\hat{\bf x}$, resulting interband superpositions between $d$-character bands contain components such as $|d_{xz}\rangle \pm i |d_{yz}\rangle=|L_z=\pm \hbar\rangle$ and $|d_{x^2-y^2}\rangle \pm i |d_{xy}\rangle=|L_z=\pm 2 \hbar \rangle$,
resulting in OHE and SHE. See {\it Discussion} for the example of fcc Pt.

For three-dimensional semiconductors such as GaAs~\cite{Murakami2003} and HgTe~\cite{Murakami2004}, roles of the momentum-space Berry curvature on SHE are analyzed near the $\Gamma$ point, at which $p$-character bands become four-fold degenerate. Since the analyses ignore the inversion symmetry breaking in the semiconductors, they apply to our model system [Fig.~\ref{fig:model}(a)] as well and explain small narrow peaks (on top of large broad background) at $E_{\rm F}\approx -0.2$ and $-0.7$ eV in Fig.~\ref{fig:Hall-conductivities}(b), which arise from the four-fold degenerate $\Gamma$ and H points [Fig.~\ref{fig:model}(b)].
Although roles of ${\bf L}$ are not evident in the analyses, 
they may also be interpreted by the orbital-texture-based mechanism except that the origin of the orbital texture is different;
in the zincblende structure of GaAs and HgTe, nearest neighbor hoppings between $p_x$, $p_y$, $p_z$ orbitals {\it themselves} can generate orbital texture in $p$-character valence bands. 
This interpretation explains OHE in hole-doped Si~\cite{Bernevig2005} naturally, which may be regarded as the vanishing SOC limit of hole-doped GaAs.
It also predicts $p$-character bands with the total spin 
$J=1/2$ and $3/2$ 
to contribute to SHC oppositely
due to their opposite correlation ${\bf L}\cdot{\bf S}$,
which is consistent with the first-principles calculation results~\cite{Guo2005,Yao2005} 
when both types of bands are partially occupied in hole-doped GaAs.

SHE in semiconductors is examined also for $s$-character conduction bands~\cite{Yao2005,Rashba2008} near the $\Gamma$ point. Related with the result, we remark that the orbital-texture-based mechanism applies even for the $s$-character band in Fig.~\ref{fig:model}(b) since the $s$-character band has partial $p$-character due to the $sp$ hopping and thus can have nonzero $\langle {\bf L} \rangle$ through the interband superpositions with $p$-character bands. To demonstrate this point, we calculate $\Omega_{nm{\bf k}}^{L_z}$ [Eq.~(\ref{eq:Kubo_1b})] near ${\bf k}={\bf 0}$ with $m$ denoting the $s$-character band and $n$ one of the $p$-character bands in the limit $\alpha_{\rm so}=0$~\cite{opposite-limit} and $\Delta_{\rm cf}=0$. Considering that $-\Omega_{nm{\bf k}}^{L_z}$ may alternatively be interpreted [Eq.~(\ref{eq:Kubo_1a})] as a contribution to OHE in $p$-character bands through their interband superposition with the $s$-character band, $\Omega_{nm{\bf k}}^{L_z}$ may be evaluated from properties of $p$-character bands. 
For the $p$ character bands, we obtain~\cite{Supplementary} the Berry curvature $\boldsymbol{\Omega}^{(\! p)} (\mathbf{k})$ arising from their interband superposition with the $s$-character band,
\begin{equation}
\label{eq:main_2}
\boldsymbol{\Omega}^{(\! p)} (\mathbf{k})
=
- 2\lambda_{sp} \mathbf{L}^{(\! p)} +\mathcal{O}({\bf k})^2,
\end{equation}
where
$\lambda_{sp}=a^2 \gamma_{sp}^2/2\hbar E_g^2$, $a$ is the lattice spacing of the cubic lattice in Fig.~2(a), $E_g$ is the band gap between $s$- and $p$-character bands, and $\mathbf{L}^{(\! p)}$ is the operator ${\bf L}$ projected to the $p$-character sub-Hilbert space.
When the space is represented by the three basis orbitals $|p_x\rangle$, $|p_y\rangle$, $|p_z\rangle$,
$\mathbf{L}^{(p)} = (L^{(p)}_x, L^{(p)}_y,L^{(p)}_z )$ is represented by the following elements
\begin{equation}
\label{eq:L for L=1}
\left(
L_\alpha^{(p)}
\right)
_{\beta\gamma} = -i{\hbar}\epsilon_{\alpha\beta\gamma}.
\end{equation}
%
The matrices satisfy the usual angular momentum commutation relations.
$\boldsymbol{\Omega}^{(\! p)}({\bf k})$ is thus {\it non-Abelian}~\cite{Murakami2003}. 
The non-Abelian nature of $\boldsymbol{\Omega}^{(\! p)}({\bf k})$ 
is a consequence of symmetries since Abelian Berry curvatures are forced to vanish~\cite{Gradhand2012} by the combination of the space inversion and the time-reversal symmetries. Only non-Abelian Berry curvatures can survive the symmetry constraints. When real wavefunctions are used as bases of representations as in Eq.~(\ref{eq:L for L=1}), the symmetries force only diagonal components of the ${\bf k}$-space Berry curvature to vanish and off-diagonal components are free from such constraints.
It is these off-diagonal components of $\boldsymbol{\Omega}^{(\! p)}({\bf k})$ that generate intrinsic OHE.
From $\boldsymbol{\Omega}^{(\! p)}({\bf k})$, one obtains~\cite{Supplementary} near ${\bf k}={\bf 0}$, 
\begin{equation}
\Omega_{nm{\bf k}}^{L_z}\approx \frac{1}{4}\!\!  \sum_{\mu,\nu=x,y,z} \!\!
\textup{Re}
\left[\langle u_{n{\bf k}}|p_\mu\rangle \left(L_z^{(p)}\Omega_z^{(p)}\right)_{\mu\nu}\langle p_\nu|u_{n{\bf k}}\rangle
\right],
\end{equation}
which confirms OHE in the $s$-character band through its interband superposition with the $p$-character bands.

{\it Discussion.---} 
In addition to the $sp$ hybridized system in a simple cubic lattice [Fig.~\ref{fig:model}(a)],
we perform calculations for other orbital hybridizations in other centrosymmetric systems and obtain similar results~\cite{Jo2018}.  
For fcc Pt, in particular, we verify~\cite{Supplementary} that as the strength of the orbital texture is gradually reduced in calculation,
its SHC reduces to zero just like Fig.~\ref{fig:Hall-conductivities}(b). This indicates that 
the orbital-texture-based mechanism is the dominant mechanism of SHE in fcc Pt and that strong SOC alone is not sufficient and orbital texture is crucial.

Since orbital(spin) currents are not directly measurable, OHE(SHE) should be probed through orbital(spin) moment accumulated at edges of systems~\cite{Sinova2006}. The magneto-optical Kerr effect is used in Refs.~\cite{Kato2004, Stamm2017} to probe accumulated magnetic moments at edges. To differentiate the orbital and spin accumulations, one may utilize X-ray magnetic circular dichroism~\cite{Brien1994, Bonetti2017} or electron energy-loss magnetic circular dichroism~\cite{Schattschneider2006,Verbeeck2010}.

Since orbital(spin) is not conserved, the connection between edge accumulation and $\sigma_{\rm OH(SH)}$ is not straightforward and there is ongoing discussion~\cite{Shi2006}. To assess this connection, we calculate the non-equilibrium orbital(spin) density generated by ${\bf E}$ as a function of position for a finite size system. We verify~\cite{Supplementary} that two opposite edges have opposite signs of orbital(spin) accumulations and for a given edge, the orbital(spin) accumulations at two different $E_{\rm F}$ values ($-1.3$ and $+0.0$ $\mathrm{eV}$, respectively) have the opposite(same) signs, which agree qualitatively with the behaviors of $\sigma_{\rm OH(SH)}$ in Fig.~3. However further study is needed to clarify the connection, which goes beyond the scope of this Letter.

To conclude, we demonstrated that orbital texture in multi-orbital centrosymmetric systems can generate OHE, which is then converted to SHE by SOC. We found that SHE does not necessarily monotonically increase with SOC strength, which provides one possible explanation why experiments (see Table III in Ref.~\cite{Reynolds2017} and Ref.~\cite{Ueda2016}) on $f$ orbital rare-earth systems with very strong SOC find $\sigma_{\rm SH}$ to be only $100\sim 200\ (\hbar/2|e|)\ (\Omega\cdot{\rm cm})^{-1}$, which is about one order of magnitude smaller than that for Pt.
According to our preliminary calculation~\cite{Jo2018}, systems with much weaker SOC such as vanadium can have $\sigma_{\rm SH}\sim - 200\ (\hbar/2|e|)\ (\Omega\cdot{\rm cm})^{-1}$, which is converted from exceptionally large $\sigma_{\rm OH}\sim 10^4 \ (\hbar/2|e|)\ (\Omega\cdot{\rm cm})^{-1}$ by weak SOC.
Considering the importance of orbital hybridization, we argue that stronger OHE and SHE may be realized in binary compounds, in which orbitals of different character from different atomic elements share similar energies and generate strong orbital hybridization.
Our result calls for experimental efforts to probe dynamically generated ${\bf L}$ in materials with strong OHE or SHE.

D.G. acknowledges fruitful discussion with Wonsig Jung and Yuriy Mokrousov. We thank E. I. Rashba for valuable comments on similarity between Eq.~(\ref{eq:main_2}) and Ref.~\cite{Rashba2008}. D.G. was supported by Global Ph.D. Fellowship Program by National Research Foundation of Korea (Grant No. 2014H1A2A1019219). D.J. and H.W.L. were supported by the SSTF (Grant No. BA-1501-07). C.K.'s work on the Berry curvature was supported by the research program of Institute for Basic Science (Grant No. IBS-R009-G2).

\let\oldaddcontentsline\addcontentsline
\renewcommand{\addcontentsline}[3]{}
%
\let\addcontentsline\oldaddcontentsline

\pagebreak
\widetext

\setcounter{equation}{0}
\setcounter{figure}{0}
\setcounter{table}{0}
\setcounter{page}{1}

\renewcommand{\theequation}{S\arabic{equation}}
\renewcommand{\thefigure}{S\arabic{figure}}
\renewcommand{\bibnumfmt}[1]{[S#1]}
\renewcommand{\citenumfont}[1]{S#1}
\renewcommand{\thepage}{S\arabic{page}}  

\setlength{\textfloatsep}{25pt}
\setlength{\parskip}{0pt}

\begin{center}
	\textbf{\large \bf Supplementary Information for
		\\
		``Intrinsic Spin and Orbital Hall Effects from Orbital Texture"}
\end{center}
\begin{center}
	{ Dongwook Go, Daegeun Jo, Changyoung Kim, and Hyun-Woo Lee$^{*}$}
\end{center}

\tableofcontents

\subsection{ A. \ \ \ Three-dimensional generalization of Fig.~1(a)}

A generalization of Fig.~1(a) in three dimensions requires one to include $p_z$ orbitals as well as $p_x$ and $p_y$ orbitals. The main difference compared to the two-dimensional case [Fig.~1(a)] is that there are two types of tangential orbitals. In Figs. S1(a) and S1(b), radial and tangential orbitals are schematically shown at the Fermi surface, respectively.

In Figs.~1(a) and 1(b), we completely neglected the $p_z$ orbital and considered only  the $p_x$ and $p_y$ orbitals (to be precise, one type of radial orbitals and one type of tangential orbitals, which are superpositions of the $p_x$ and $p_y$ orbitals). Such neglect of the $p_z$ orbital is possible in the $k_z=0$ plane since the $k_z=0$ plane is a mirror reflection symmetry plane where, the $p_z$ orbital (odd under the mirror reflection) does not hybridize with the $p_x$ or $p_y$ orbitals (even under the mirror reflection) due to their parity difference.

\begin{figure}[b]
	\label{fig:orbital_character_3D}
	\includegraphics[angle=0, width=0.5\textwidth]{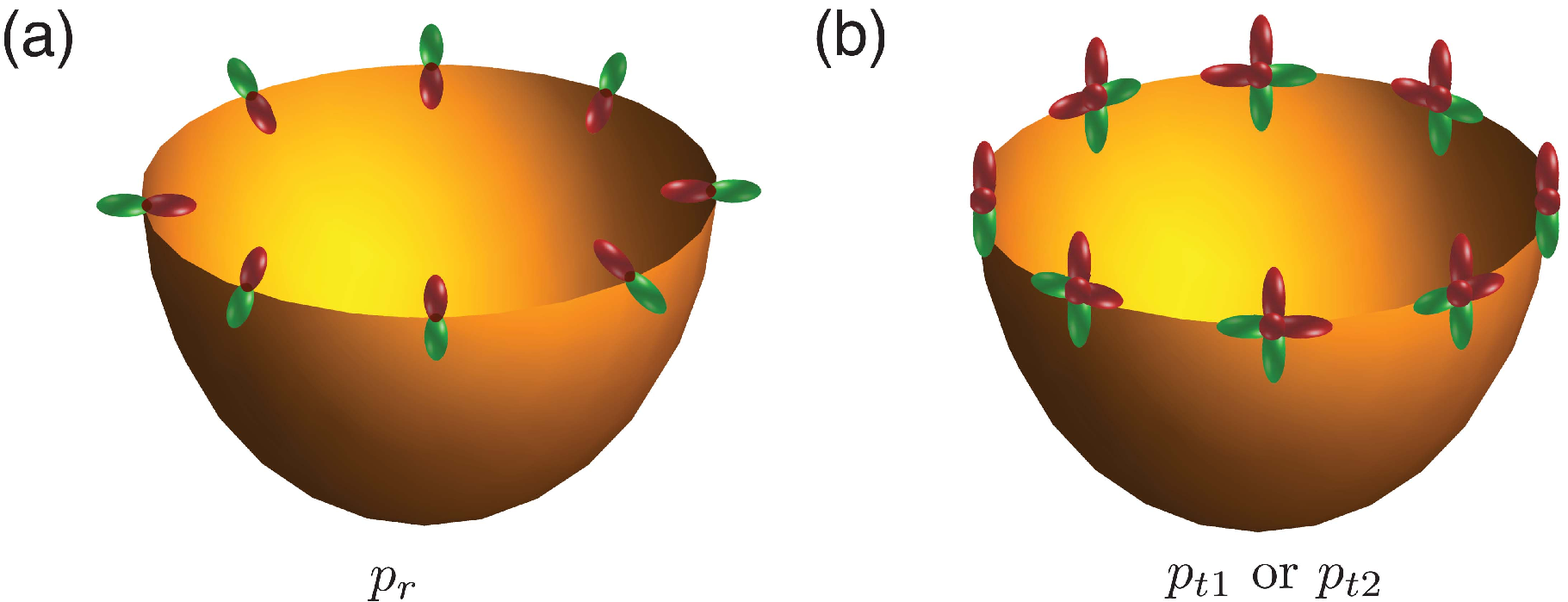}
	\caption{ 
		(a) Radial ($p_r$) and (b) tangential ($p_{t1}$ or $p_{t2}$) orbitals on the Fermi surface in a three dimensional model including $p_x$, $p_y$, $p_z$ orbitals. Note that there are two types of tangential orbitals. }
\end{figure}

\subsection{B. \ \ \ ${sp}$ model in the simple cubic lattice}

Detailed information on the $sp$ model used for the numerical calculation in the Letter is presented in this section. We assume that there are $s$, $p_x$, $p_y$, $p_z$ orbitals $\ket{\phi_{n\sigma\mathbf{R}}}$ at each Bravais lattice in the simple cubic structure. By using Bloch-like states
\begin{eqnarray}
	\label{eq:Bloch-like_state}
	e^{i\mathbf{k}\cdot\mathbf{r}}\ket{\varphi_{n\sigma\mathbf{k}}}
	=
	\sum_\mathbf{R}
	e^{i\mathbf{k}\cdot\mathbf{R}}
	\ket{\phi_{n\sigma\mathbf{R}}},
\end{eqnarray}
as basis, where $n=s, p_x, p_y, p_z$ is the orbital character and $\sigma=\uparrow,\downarrow$ is the spin, the tight-binding Hamiltonian in $\mathbf{k}$-space is written as
\begin{eqnarray}
	H_\textup{tot} (\mathbf{k})
	=
	H_\textup{kin} (\mathbf{k})
	+
	H_\textup{so},
\end{eqnarray}
where $H_\textup{kin} (\mathbf{k})$ is the kinetic part of the Hamiltonian arising from hoppings and on-site energies, and $H_\textup{so}$ describes spin-orbit coupling (SOC) near the atomic nuclei.
First, $H_\textup{kin} (\mathbf{k})$ is independent of the spin and its nonzero matrix elements are
\begin{eqnarray}
	& &
	\bra{\varphi_{s\sigma\mathbf{k}}}
	H_\textup{kin}
	\ket{\varphi_{s\sigma\mathbf{k}}}
	=
	E_s
	-2 t_s
	\left[
	\cos (k_x a)
	+
	\cos (k_y a)
	+
	\cos (k_z a)
	\right],
	\\
	& &
	\bra{\varphi_{p_x\sigma\mathbf{k}}}
	H_\textup{kin}
	\ket{\varphi_{p_x\sigma\mathbf{k}}}
	=
	E_{p_x}
	+
	2 t_{p\sigma}
	\cos (k_x a)
	-2t_{p\pi}
	\left[
	\cos (k_y a)
	+
	\cos (k_z a)
	\right],
	\\
	& &
	\bra{\varphi_{p_y\sigma\mathbf{k}}}
	H_\textup{kin}
	\ket{\varphi_{p_y\sigma\mathbf{k}}}
	=
	E_{p_y}
	+
	2 t_{p\sigma}
	\cos (k_y a)
	-2t_{p\pi}
	\left[
	\cos (k_z a)
	+
	\cos (k_x a)
	\right],
	\\
	& &
	\bra{\varphi_{p_z\sigma\mathbf{k}}}
	H_\textup{kin}
	\ket{\varphi_{p_z\sigma\mathbf{k}}}
	=
	E_{p_z}
	+
	2 t_{p\sigma}
	\cos (k_z a)
	-2t_{p\pi}
	\left[
	\cos (k_x a)
	+
	\cos (k_y a)
	\right],
	\\
	& &
	\bra{\varphi_{s\sigma\mathbf{k}}}
	H_\textup{kin}
	\ket{\varphi_{p_x\sigma\mathbf{k}}}
	=
	2i\gamma_{sp}
	\sin (k_x a),
	\\
	& &
	\bra{\varphi_{s\sigma\mathbf{k}}}
	H_\textup{kin}
	\ket{\varphi_{p_y\sigma\mathbf{k}}}
	=
	2i\gamma_{sp}
	\sin (k_y a),
	\\
	& &
	\bra{\varphi_{s\sigma\mathbf{k}}}
	H_\textup{kin}
	\ket{\varphi_{p_z\sigma\mathbf{k}}}
	=
	2i\gamma_{sp}
	\sin (k_z a).
\end{eqnarray}
Here, $E_s$ and $E_{p_i}$ are on-site energies for $s$ and $p_i\ (i=x,y,z)$ orbitals, respectively, and $t_s$, $t_{p\sigma(\pi)}$, $\gamma_{sp}$ are the nearest neighbor hopping amplitudes between $s$ orbitals, between $p$ orbitals via $\sigma(\pi)$ bonding, and between $s$ and $p$ orbitals, respectively.
Second, SOC is
\begin{eqnarray}
	H_\textup{so}
	=
	\frac{2\alpha_\textup{so}}{\hbar^2}
	\mathbf{L}
	\cdot
	\mathbf{S},
\end{eqnarray}
where $\mathbf{L}(\mathbf{S})$ is the orbital(spin) angular momentum operator near atomic nuclei. For $p$ orbitals, $\mathbf{L}=(L_x, L_y, L_z)$ becomes
\begin{eqnarray}
	L_x = \hbar
	\begin{pmatrix}
		0 & 0 & 0 \\
		0 & 0 & -i \\
		0 & i & 0
	\end{pmatrix},
	\
	L_y = \hbar
	\begin{pmatrix}
		0 & 0 & i \\
		0 & 0 & 0 \\
		-i & 0 & 0
	\end{pmatrix},
	\
	L_x = \hbar
	\begin{pmatrix}
		0 & -i & 0 \\
		i & 0 & 0 \\
		0 & 0 & 0
	\end{pmatrix},
\end{eqnarray}
in the matrix representation using the basis states $\varphi_{p_x\sigma\mathbf{k}}$, $\varphi_{p_y\sigma\mathbf{k}}$, and $\varphi_{p_z\sigma\mathbf{k}}$. The spin angular momentum opeators are represented by the Pauli matrices within each orbital:
\begin{eqnarray}
	\bra{\varphi_{n\sigma\mathbf{k}}}
	\mathbf{S}
	\ket{\varphi_{n\sigma'\mathbf{k}}}
	=
	\frac{\hbar}{2}
	\left[
	\boldsymbol{\sigma}
	\right]_{\sigma\sigma'}
	.
\end{eqnarray}
Values of the parameters used in the calculation are
$
E_s = 3.2,\
E_p = E_{p_x} = E_{p_y} = E_{p_z} = -0.5, \
t_s = 0.5, \
t_{p\sigma} = 0.5, \
t_{p\pi} = 0.2, \
\gamma_{sp} = 0.5, \
\alpha_\textup{so} = 0.1,
$
all in unit of $\mathrm{eV}$. All parameters are set as above unless specified otherwise, such as in Fig. 3.

\subsection{C. \ \ \ SOC dependences of OHC and SHC}

\begin{figure}[t]
	\label{fig:SOC_dependence}
	\includegraphics[angle=0, width=0.45\textwidth]{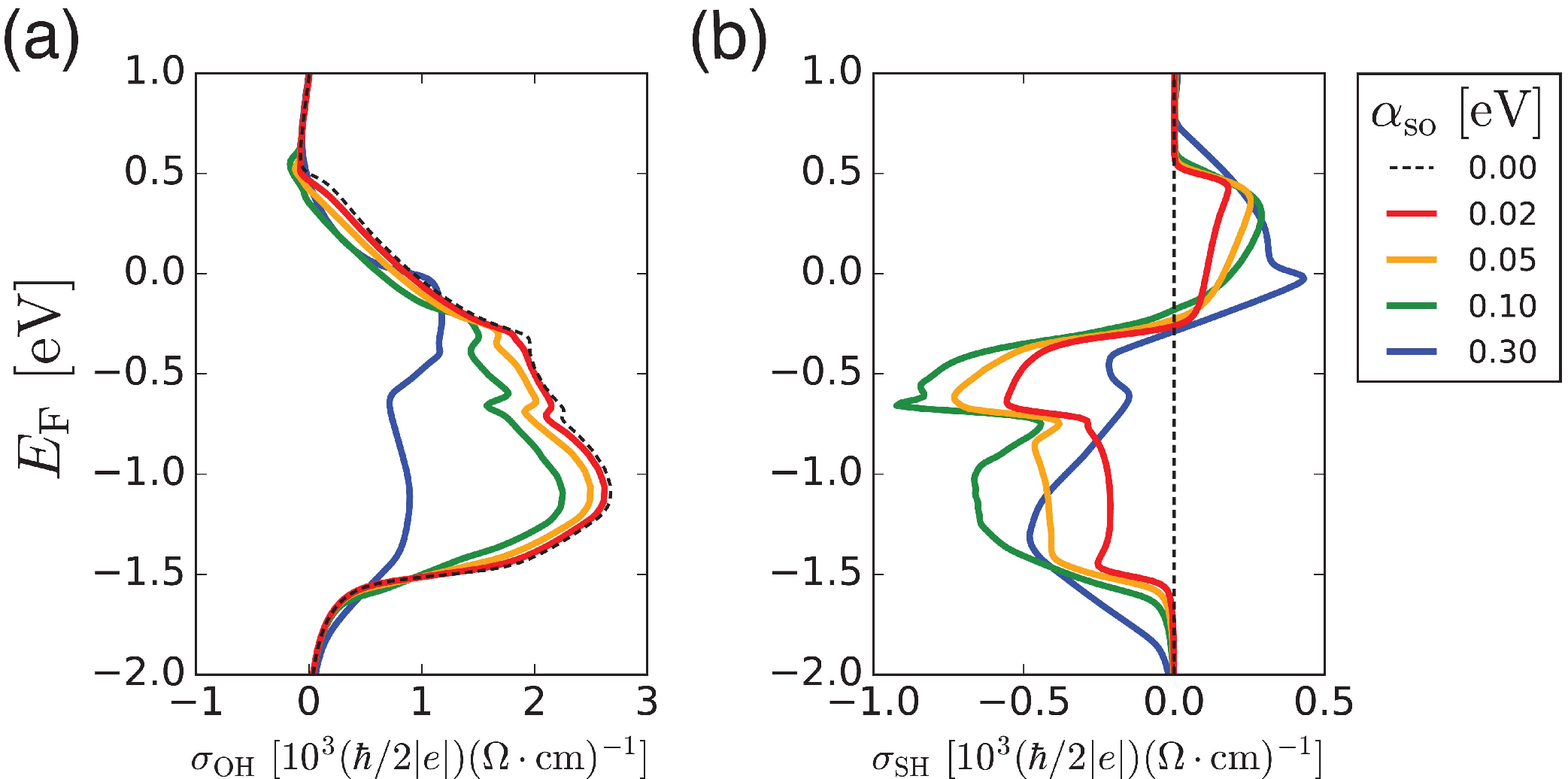}
	\caption{
		SOC dependences of (a) OHC ($\sigma_\mathrm{OH}$) and (b) SHC ($\sigma_\mathrm{SH}$) as a function of Fermi energy $E_\mathrm{F}$. Note that $\sigma_\mathrm{SH}$ exhibits non-monotonic behavior with different SOC strength $\alpha_\mathrm{so}$.
	}
\end{figure}

In the Letter, the SOC dependences of OHC and SHC are shown only for a fixed electron density. Here, we present OHC and SHC as a function of Fermi energy for different values of the SOC strength $\alpha_\mathrm{so}$ [Fig. S2(a) and S2(b)]. It is observed that while OHC monotonically decreases with $\alpha_\mathrm{so}$, SHC exhibits nonmonotonic behavior for a wide range of $E_\mathrm{F}$.

\subsection{D. \ \ \ $\mathbf{k}$- and band-resolved contributions of OHC and SHC}

\begin{figure}[t]
	\label{fig:kres_contribution}
	\includegraphics[angle=0, width=0.45\textwidth]{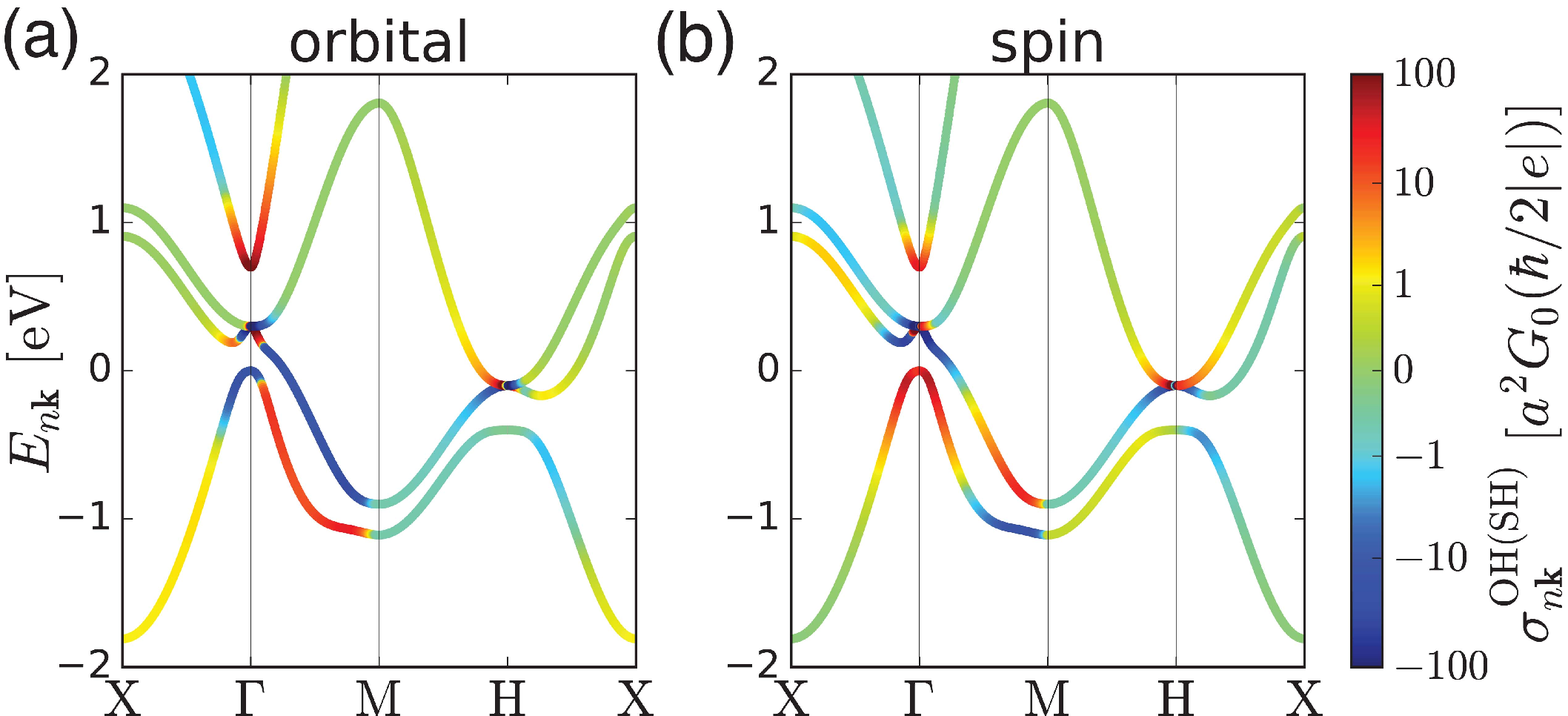}
	\caption{
		Plots for (a) $\sigma_{n\mathbf{k}}^\mathrm{OH}$ and (b) $\sigma_{n\mathbf{k}}^\mathrm{SH}$ (shown in color) on top of the band structure. Near the high-symmetry points, such as $\Gamma$ and $\mathrm{H}$, we find that the sign of the SHC is the same(opposite) to that of the OHC if the correlation $\left\langle \mathbf{L}\cdot\mathbf{S}\right\rangle$ is positive(negative). This qualitatively explains the correlation between the OHC and SHC induced by the SOC.
	}
\end{figure}

We argued in the Letter that spin Hall effect (SHE) is a secondary effect and converted from orbital Hall effect (OHE) by SOC. 
Each state $\ket{u_{n\mathbf{k}}}$ tends to contribute to SHC with same(opposite) sign to the sign of the corresponding contribution to OHC if the correlation $\left\langle \mathbf{L}\cdot\mathbf{S}\right\rangle$ [Fig. 2(b)] is positive(negative). In order to compare each contribution in the band structure, we define $\mathbf{k}$- and band-resolved contribution $\sigma_{n\mathbf{k}}^\mathrm{OH(SH)}$ as
\begin{align}
	\sigma_\mathrm{OH(SH)} = \sum_n \int \frac{d^3k}{(2\pi)^3} f_{n\mathbf{k}} \sigma_{n\mathbf{k}}^\mathrm{OH(SH)}.
\end{align}
By comparing Eq.~(S13) with Eq. (2) in the Letter, we find that
\begin{align}
	\sigma_{n\mathbf{k}}^\mathrm{OH(SH)} = 
	-
	2e\hbar 
	\sum_{m \neq n}
	\textup{Im}
	\left[
	\frac{
		\bra{u_{n\mathbf{k}}}
		j_y^{X_z}
		\ket{u_{m\mathbf{k}}}
		\bra{u_{m\mathbf{k}}}
		v_x
		\ket{u_{n\mathbf{k}}}
	}
	{(E_{n\mathbf{k}} - E_{m\mathbf{k}}+i\eta)^2}
	\right],
\end{align}
where $j_y^{X_z} = (v_y X_z + X_z v_y)/2$ is the conventional orbital(spin) current with $X_z = L_z(S_z)$. In Fig. S3(a) and S3(b), $\sigma_{n\mathbf{k}}^\mathrm{OH}$ and $\sigma_{n\mathbf{k}}^\mathrm{SH}$ are shown in color on top of the band structure, respectively. It can be seen that their relative sign profile qualitatively match with that of $\left\langle \mathbf{L}\cdot\mathbf{S} \right\rangle$, which is especially clear near the high symmetry points such as $\Gamma$ and $\mathrm{H}$, although there are quantitative deviations in sign-changing positions of $\sigma_{n\mathbf{k}}^\mathrm{OH(SH)}$.

\subsection{E. \ \ \ Stability of OHC and SHC against the crystal field splitting}

\begin{figure}[b]
	\includegraphics[angle=0, width=0.45\textwidth]{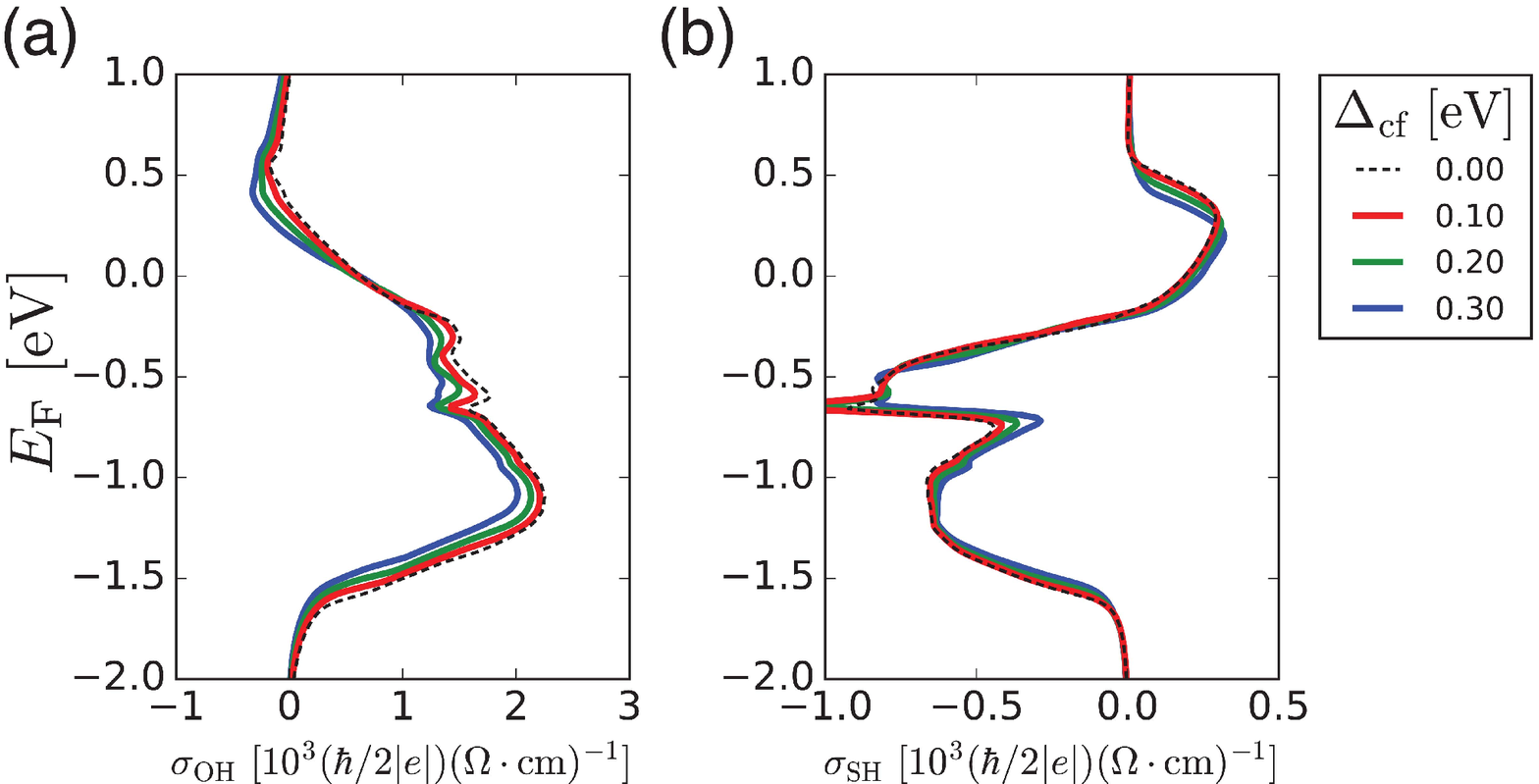}
	\caption{
		Fermi energy ($E_\mathrm{F}$) dependences of (a) OHC ($\sigma_\mathrm{OH}$) and (b) SHC ($\sigma_\mathrm{SH}$) for different values of crystal field splitting $\Delta_\mathrm{cf}$. Note that both $\sigma_\mathrm{OH}$ and $\sigma_\mathrm{SH}$ remain stable against $\Delta_\mathrm{cf}$.
	}
\end{figure}

In order to check stability of OHE and SHE against the crystal field splitting, we calculate OHC and SHC by changing on-site energies of $p$ orbitals as
\begin{eqnarray}
	E_{p_x} = E_p + \Delta_\mathrm{cf},\ E_{p_y} = E_p - \Delta_\mathrm{cf},\ E_{p_z} = E_p.
\end{eqnarray}
In Figs.~S4(a) and S4(b), Fermi energy dependences of OHC and SHC are shown for different values of $\Delta_\mathrm{cf}$. It can be seen that $\Delta_\mathrm{cf}$ has a negligible effect on OHC and SHC despite the fact that the square of the energy difference appears in the denominator of the Kubo formula in Eq.~(2b). The reason for the insensitivity is that even for $\Delta_{\rm cf}=0$, there occurs energy splitting between different $p$ character bands due to the $sp$ hybridization $\gamma_{sp}$ [Fig.~2(b)]. Thus in a situation where the energy splitting due to the $sp$ hybridization is sufficiently larger than $\Delta_{\rm cf}$, $\Delta_{\rm cf}$ cannot affect the OHC or SHC significantly. Note that the $sp$ hybridization strength $\gamma_{sp}$ is set to 0.5 eV (except when $\gamma_{sp}$ is varied in Figs.~3(a) and 3(b)) whereas $\Delta_{\rm cf}$ is varied only up to 0.3 eV, thus satisfying the condition $\gamma_{sp} \gtrsim \Delta_\mathrm{cf}$. To demonstrate this point further, we calculate Fermi energy dependences of OHC and SHC for much smaller value of $sp$ hybridization, $\gamma_{sp}=0.05\ \mathrm{eV}$ [Fig.~S5(a) and S5(b)]. In this situation, the OHC and SHC are suppressed rather significantly with $\Delta_{\rm cf}$.

To be more rigorous, there are a few exceptional ${\bf k}$-points at which $\gamma_{sp}$ cannot generate an energy splitting. They are high symmetry points such as $\Gamma$ and $\mathrm{H}$ in case of the simple cubic lattice addressed in the Letter. Contributions from those special ${\bf k}$ point to the OHC and SHC are sensitive to $\Delta_{\rm cf}$ since the square of the energy difference appears in Eq.~(2b) and the difference is zero for $\Delta_\mathrm{cf}=0$. For $\Delta_{\rm cf}=0$, in particular, $\Omega_{nm{\bf k}}^{X_z}$ in Eq.~(2b) become divergent $\sim|{\bf k}-{\bf k}_{\Gamma({\rm H})}|^{-2}$ near the special $\mathbf{k}$-points. This divergence is the same kind of divergence as the divergent Berry curvature in $p$-doped semiconductors reported by Murakami \emph{et al.} \cite{Murakami2003_supp}. However, three-dimensional integral of $\Omega_{nm{\bf k}}^{X_z}$ over a small volume near ${\bf k}_{\Gamma({\rm H})}$ is finite and {\it not} divergent. 
According to our calculation in the simulation part, the only symptom of the divergent $\Omega_{nm{\bf k}}^{X_z}$ is the emergence of small and narrow peaks or dips in the $E_{\rm F}$ dependence of the OHC [Fig.~3(a)] and SHC [Fig.~3(b)]. This explains why the $E_{\rm F}$ dependence of the SHC exhibits a small and narrow peak at $E_{\rm F}=-0.7$ eV (which corresponds to a quadratic band touching at $\mathrm{H}$) and this peak is relatively sensitive to $\Delta_{\rm cf}$. Note however that even for those special $E_{\rm F}$ values, the $\Delta_{\rm cf}$ dependence is not so drastic since the ${\bf k}$-integration over the divergent integrand $\Omega_{nm{\bf k}}^{X_z}$ produces a {\it nondivergent} result and other ${\bf k}$-points away from those exceptional ${\bf k}$-points contribute sizably to the OHC and SHC. In line with this observation, Tanaka {\it et al.} reported that the main contribution to the intrinsic SHC in Pt arises from a rather wide range of ${\bf k}$-points away from those exceptional high symmetry points \cite{Tanaka2008_supp}.

\begin{figure}[t]
	\includegraphics[angle=0, width=0.5\textwidth]{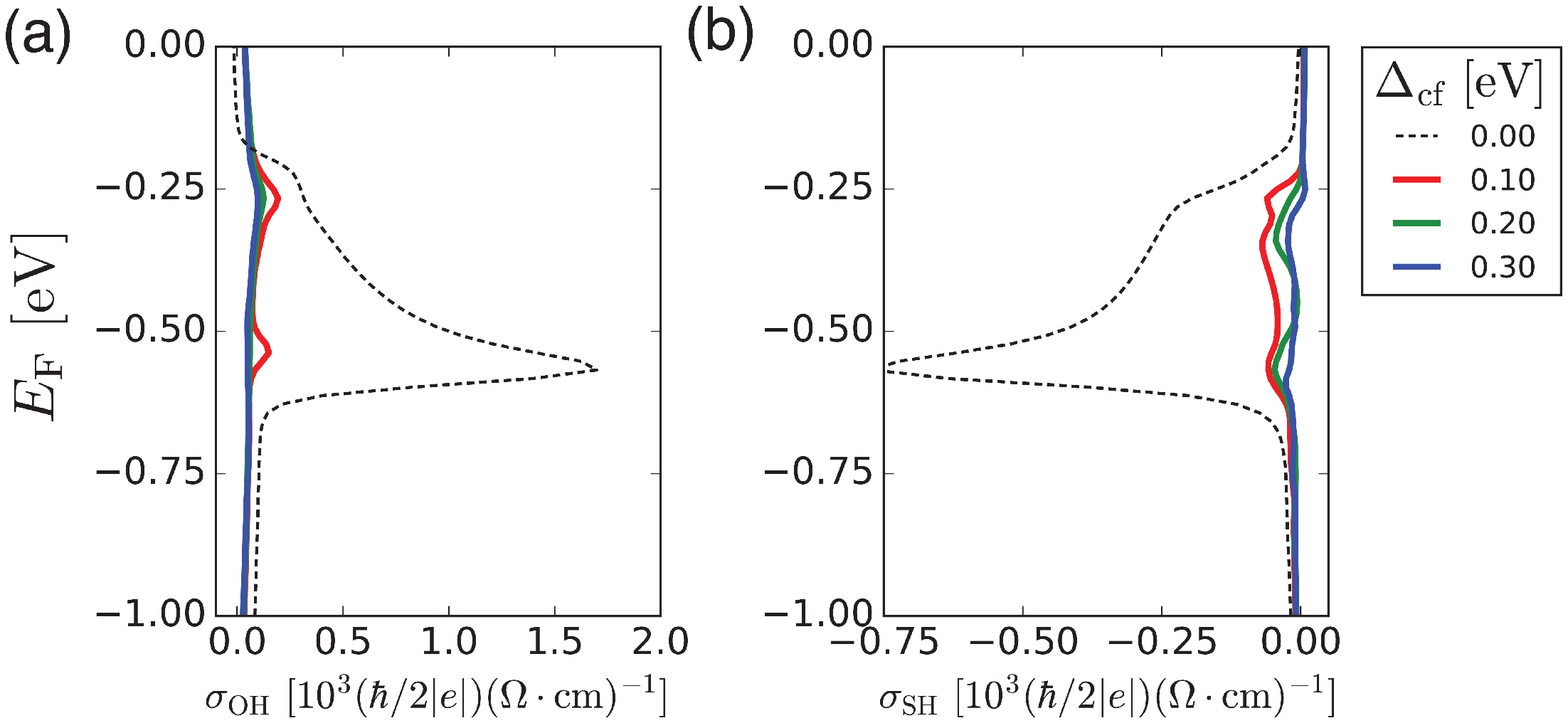}
	\caption{
		Fermi energy ($E_\mathrm{F}$) dependences of (a) OHC ($\sigma_\mathrm{OH}$) and (b) SHC ($\sigma_\mathrm{SH}$) for different values of crystal field splitting $\Delta_\mathrm{cf}$ for smaller value of $sp$ hybridization $\gamma_{sp}=0.05\ \mathrm{eV}$. Compared to Fig. S4 (where $\gamma_{sp}=0.50\ \mathrm{eV}$), both $\sigma_\mathrm{OH}$ and $\sigma_\mathrm{SH}$ are significantly suppressed by $\Delta_\mathrm{cf}$.
	}
\end{figure}

\subsection{F. \ \ \ \ Non-Abelian Berry curvature from the $sp$ hybridization}

In this section, we calculate the non-Abelian Berry curvature in Eq. (3). We use the quasi-degenerate perturbation theory explained in Refs. \cite{Nozieres1973_supp, Winkler_textbook_supp}. In the $sp$ model, the Hamiltonian in $\mathbf{k}$-space is formally written as
\begin{align}
	H (\mathbf{k})
	=
	\begin{pmatrix}
		H_p (\mathbf{k}) & h(\mathbf{k}) \\ 
		h^\dagger (\mathbf{k}) & H_s (\mathbf{k})
	\end{pmatrix},
\end{align}
where $H_{p(s)} (\mathbf{k})$ is the Hamiltonian within a subspace spanned by $p(s)$ orbitals, and $h(\mathbf{k})$ describes hopping between a $s$-character band and $p$-character bands. The Berry phase effect becomes manifest by projecting the dynamics in the ground state within the adiabatic approximation. This can be achieved by applying a unitary operator $U(\mathbf{k})$ which eliminates $h(\mathbf{k})$:
\begin{align}
	H_\textup{eff} (\mathbf{k})
	=
	U(\mathbf{k}) H (\mathbf{k}) U^\dagger (\mathbf{k})
	=
	\begin{pmatrix}
		H_{p,\textup{eff}} (\mathbf{k}) & 0 \\ 
		0 & H_{s,\textup{eff}} (\mathbf{k})
	\end{pmatrix}.
\end{align}
At the same time, all the observable operators transform in the same way. Importantly, the canonical position operator transforms as
\begin{align}
	\mathbf{r} \mapsto \mathbf{R} = U(\mathbf{k}) \mathbf{r} U^\dagger (\mathbf{k}) = \mathbf{r} + \boldsymbol{\mathcal{A}} (\mathbf{k}),
\end{align}
where $\boldsymbol{\mathcal{A}} (\mathbf{k}) = i U(\mathbf{k}) \partial_\mathbf{k} U^\dagger (\mathbf{k})$ is the Berry connection arising from $h(\mathbf{k})$. In a perturbative regime where $| h(\mathbf{k}) | \ll E_g$, where $E_g = E_p (\mathbf{k}) - E_s (\mathbf{k})$ is the energy gap between $s$ and $p$ bands, the Berry connection in the $p$ orbital subspace is written as
\begin{align}
	\boldsymbol{\mathcal{A}}^{(p)} (\mathbf{k})
	=
	\frac{i}{2}
	\left[
	h(\mathbf{k}) 
	\frac{1}{E_g^2}
	\left\{
	\partial_\mathbf{k}
	h^\dagger (\mathbf{k}) 
	\right\}
	-
	\left\{
	\partial_\mathbf{k}
	h (\mathbf{k}) 
	\right\}
	\frac{1}{E_g^2}
	h^\dagger (\mathbf{k})
	\right].
\end{align}
The assumption above always holds near the high-symmetry points since 
\begin{align}
	h (\mathbf{k})
	=
	-2i\gamma_{sp}
	\left[
	\sin (k_x a),
	\
	\sin (k_y a),
	\
	\sin (k_z a)
	\right]^{\mathrm{T}}.
\end{align}
Similarly, the Berry connection in the $s$ orbital subspace can be found by interchanging $h(\mathbf{k})$ and $h^\dagger (\mathbf{k})$.

In the absence of the SOC, the Berry connection near the $\Gamma$-point in the $p$-character bands is found as 
\begin{align}
	\boldsymbol{\mathcal{A}}^{(p)} (\mathbf{k})
	=
	\lambda_{sp} \mathbf{k} \times \mathbf{L}^{(p)},
\end{align}
where $\lambda_{sp} = a^2\gamma_{sp}^2/2\hbar E_g^2$. Here, $E_g$ is evaluated at the $\Gamma$-point. On the other hand, the Berry connection in the $s$-character band is zero because there is no internal degree of freedom in the absence of SOC. The corresponding Berry curvature for the $p$-character bands is 
\begin{align}
	\Omega_\gamma^{(p)} (\mathbf{k}) = \frac{1}{2}\epsilon_{\alpha\beta\gamma} \mathcal{F}_{\alpha\beta}^{(p)} (\mathbf{k}),
\end{align}
where 
\begin{align}
	\mathcal{F}_{\alpha\beta}^{(p)} (\mathbf{k})
	=
	\partial_{k_\alpha} \mathcal{A}_\beta^{(p)} (\mathbf{k})
	-
	\partial_{k_\beta} \mathcal{A}_\alpha^{(p)} (\mathbf{k})
	- i 
	\left[
	\mathcal{A}_\alpha^{(p)} (\mathbf{k}), \mathcal{A}_\beta^{(p)} (\mathbf{k})
	\right].
\end{align}
This leads to Eq. (3) in the Letter.

In the presence of the SOC, $p$-character bands split into $j=3/2$ and $j=1/2$ multiplets. By downfolding the Hamiltonian in each multiplet, we find that the Berry connection becomes
\begin{align}
	\boldsymbol{\mathcal{A}}^{(j)} (\mathbf{k})
	=
	\lambda_{sp}^{(j)} \mathbf{k} \times \mathbf{J}^{(j)},
\end{align}
where $\mathbf{J}=\mathbf{L}+\mathbf{S}$ is the total angular momentum, and the superscript $j$ represents its projection to $j=3/2$ or $j=1/2$ multiplet. The proportionality constant slightly changes from the $\lambda_{sp}$ obtained in the absence of the SOC:
\begin{align}
	\lambda_{sp}^{(j)}
	=
	\lambda_{sp} \frac{E_g^2}{\left[E_g + \alpha_\textup{so}\left\{ j(j+1)-11/4 \right\} \right]^2}.
\end{align}
Another important consequence of the SOC is that it gives rise to the nonzero Berry connection in $s$ band as well. By applying the same procedure as above, it is found as
\begin{align}
	\boldsymbol{\mathcal{A}}^{(s)} = 
	\lambda_{sp}^{(s)} \mathbf{S}\times\mathbf{k},
\end{align}
where
\begin{align}
	\lambda_{sp}^{(s)} 
	=
	\frac{2}{3}
	(\lambda_{sp}^{j=3/2} - \lambda_{sp}^{j=1/2}),
\end{align}
thus it vanishes in the absence of the SOC. For this limit of strong SOC, similar result was also found from the Kane model for semiconductors \cite{Nozieres1973_supp, Winkler_textbook_supp, Chang2008_supp, Rashba2008_supp}.

\subsection{G. \ \ \ Relation between the Berry curvature and OHC}

In this section, we demonstrate a relation between the non-Abelian Berry curvature described in the previous section and the OHC near the high-symmetry point such as $\Gamma$, which is shown in Eq. (5) in the Letter. Near the $\Gamma$-point, Bloch states have almost pure $s$- and $p$-characters, and the velocity operator is written as
\begin{align}
	v_{x(y)}
	\approx
	\frac{2i\gamma_{sp}a}{\hbar} 
	\Big( 
	\ket{\varphi_{p_{x(y)}\mathbf{k}}} \bra{\varphi_{s\mathbf{k}}} 
	-
	\ket{\varphi_{s\mathbf{k}}}\bra{\varphi_{p_{x(y)}\mathbf{k}}}  \Big).
\end{align}
Then we expand
\begin{align}
	\Omega_{nm\mathbf{k}}^{L_z}
	&=
	\hbar^2
	\textup{Im}
	\left[
	\frac{
		\bra{u_{n\mathbf{k}}}
		j_y^{L_z}
		\ket{u_{m\mathbf{k}}}
		\bra{u_{m\mathbf{k}}}
		v_x
		\ket{u_{n\mathbf{k}}}
	}
	{(E_{n\mathbf{k}} - E_{m\mathbf{k}}+i\eta)^2}
	\right]
\end{align}
in the lowest order in $\mathbf{k}$ for the band indices $n$ and $m$ denoting one of the $p$-character bands and the $s$-character band, respectively. Note that the $v_{x(y)}$ couples $s$- and $p$-character bands, while $L_z$ couples two $p$-character bands. Thus, it follows that
\begin{align}
	\Omega_{nm\mathbf{k}}^{L_z}
	& \approx
	\frac{\hbar^2}{2}
	\sum_{l \in p_x, p_y, p_z}
	\textup{Im}
	\left[
	\frac{
		\bra{u_{n\mathbf{k}}}
		L_z
		\ket{u_{l\mathbf{k}}}
		\bra{u_{l\mathbf{k}}}
		v_y
		\ket{u_{m\mathbf{k}}}
		\bra{u_{m\mathbf{k}}}
		v_x
		\ket{u_{n\mathbf{k}}}
	}
	{(E_{n\mathbf{k}} - E_{m\mathbf{k}}+i\eta)^2}
	\right].
\end{align}
Also, the energy eigenvalues for all $p$-character bands are the same at the $\Gamma$-point, then
\begin{align}
	\Omega_{nm\mathbf{k}}^{L_z}
	& \approx
	\frac{\hbar^2}{2}
	\sum_{l \in p_x, p_y, p_z}
	\textup{Im}
	\left[
	\frac{
		\bra{u_{n\mathbf{k}}}
		L_z
		\ket{u_{l\mathbf{k}}}
		\bra{u_{l\mathbf{k}}}
		v_y
		\ket{u_{m\mathbf{k}}}
		\bra{u_{m\mathbf{k}}}
		v_x
		\ket{u_{n\mathbf{k}}}
	}
	{
		(E_{l\mathbf{k}} - E_{m\mathbf{k}} + i\eta)
		(E_{n\mathbf{k}} - E_{m\mathbf{k}} + i\eta)}
	\right]
	\\
	& =
	\frac{1}{2}
	\sum_{l \in p_x, p_y, p_z}
	\textup{Im}
	\left[
	\bra{u_{n\mathbf{k}}}
	L_z
	\ket{u_{l\mathbf{k}}}
	\braket{\partial_{k_y} u_{l\mathbf{k}}
		| u_{m\mathbf{k}}}
	\braket{ u_{m\mathbf{k}}
		| \partial_{k_x} u_{n\mathbf{k}}}
	\right]
	\\
	&=
	\frac{1}{4}
	\sum_{l \in p_x, p_y, p_z}
	\textup{Re}
	\left[
	\bra{u_{n\mathbf{k}}}
	L_z
	\ket{u_{l\mathbf{k}}}
	\Omega_{z,ln}^{(p)} (\mathbf{k})
	\right].
\end{align}
This proves Eq. (5) in the Letter.

\subsection{H. \ \ \ Orbital-texture-based mechanism in fcc Pt}

To investigate whether orbital-texture-based mechanism presented in the Letter is a dominant mechanism of the intrinsic SHE in fcc Pt, we have carried out numerical calculations for fcc Pt via the tight-binding approach. For this calculation, we have used the tight-binding parameters from Ref.~\cite{band_structure_handbook_supp}. The red lines in Fig.~S6(a) show the resulting electronic band structure, which is in good agreement with a first-principles calculation result~\cite{Guo2008_supp}. The red line in Fig.~S6(b), on the other hand, show the calculated SHC $\sigma_{\rm SH}$ as a function of the Fermi energy $E_{\rm F}$ (real value of $E_{\rm F}$ in Pt amounts to zero in this figure). This result is also in
good agreement with the first-principles calculation result~\cite{Guo2008_supp}. For the real value of $E_{\rm F}$, which is zero, this calculation predicts $\sigma_{\rm SH}\approx 2\times 10^3\, (\hbar/2|e|)\ (\Omega\cdot{\rm cm})^{-1}$, which is comparable to experimentally measured values~\cite{Kimura2007_supp, Sagasta2016_supp}. We thus believe that this tight-binding approach works as a reliable method to address the intrinsic SHC in Pt.


\ \ \ \ In order to assess how important the orbital-texture-based mechanism is for $\sigma_{\rm SH}$ in fcc Pt, we have calculated $\sigma_{\rm SH}$ as a function of the orbital texture strength [Fig.~S6(b)]. Here, 100\% refers to the calculation result for the real strength of the orbital texture and 0\% to the calculation result for an artificial situation where the orbital texture is ``completely suppressed''. The band structure for the 0\% orbital texture is shown in Fig.~S6(a). As the orbital texture is suppressed from their full strength ($100$\%) to zero ($0$\%), $\sigma_{\rm SH}$ decreases monotonically and vanishes eventually. This implies that the orbital texture is crucial for $\sigma_{\rm SH}$. Therefore, we conclude that the orbital-texture-based mechanism gives a dominant contribution in Pt.

\ \ \ \ To provide further details of the calculation, we have modulated the orbital texture strength as follows. The on-site energy parameters are not varied during the orbital texture strength variation. Also nearest and next-nearest hopping parameters between the same orbitals are not varied. This type of equal-orbital hoppings include
$6s$-$6s$ hopping, $6p_i$-$6p_i$ hopping ($i=x$, $y$, $z$), $5d_i$-$5d_i$ hopping ($i=xy$, $yz$, $zx$, $x^2-y^2$, $z^2-r^2/3$). The rest of hopping parameters that describe hoppings between {\it different} orbitals
are varied to modulate the strength of the orbital texture. For instance, 40\% in Fig.~S6(b) means that tight-binding parameters for all those different-orbital hoppings are reduced to 40\% of their real values. Note that this kind of controllability is one advantage of tight-binding approach in contrast to first-principles calculations.

\begin{figure}[t]
	\includegraphics[angle=0, width=0.45\textwidth]{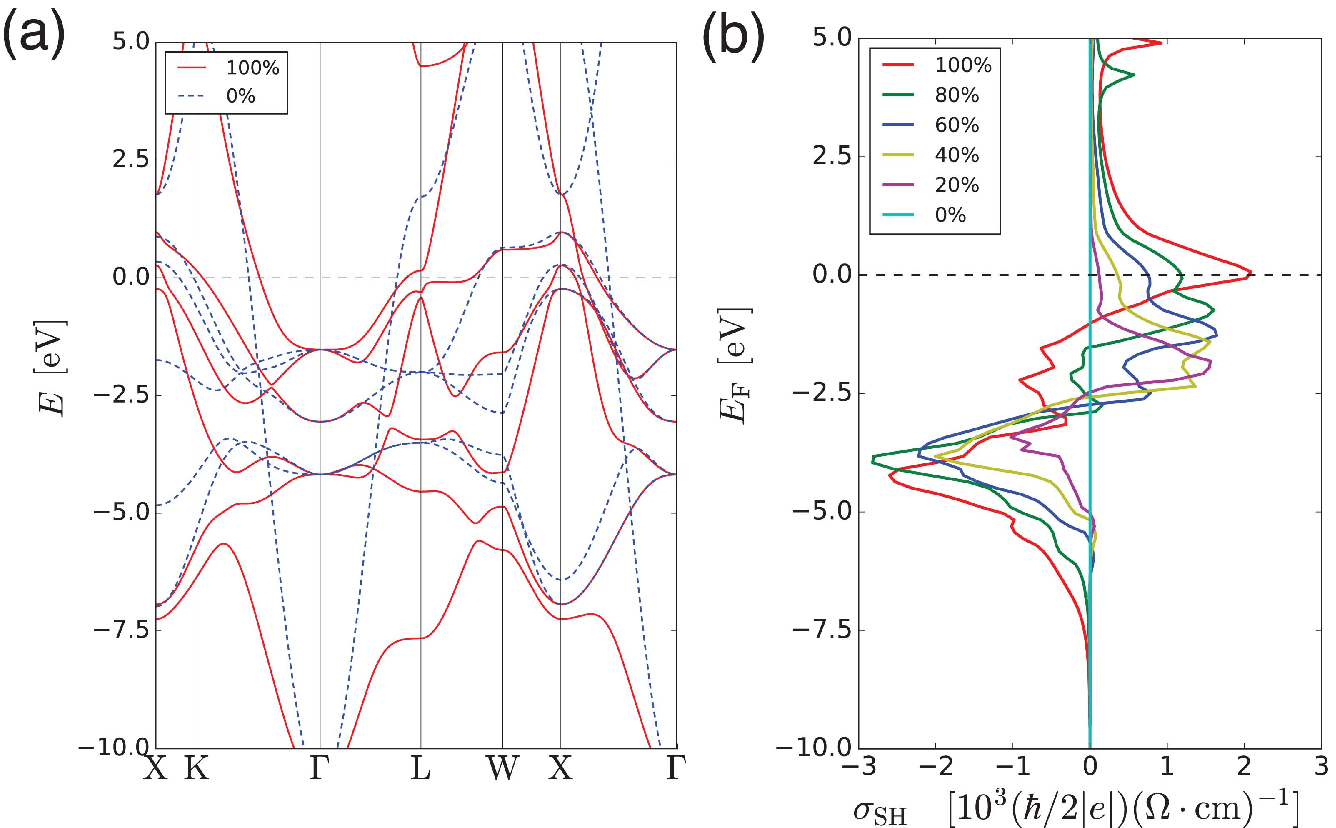}
	\caption{(a) Band structure of fcc Pt with full strength of the orbital texture (red solid line) and that with the strength of the orbital texture being artificially suppressed to zero (blue dashed line). The Fermi energy dependence of (b) SHC ($\sigma_\mathrm{SH}$) for different strengths of the orbital texture with respect to the full strength.}
\end{figure}

\ \ \  One remark is in order. The band structure of Pt in fcc structure is significantly more complicated than that of the simple cubic $sp$ system used in the Letter. Moreover, hoppings between different orbitals, which are responsible for the orbital texture, are more diverse in fcc Pt than in the simple cubic $sp$ system. 
%
Nevertheless we emphasize that both fcc Pt and the simple cubic $sp$ model exhibit similar qualitative behaviors. For instance, $\sigma_{\rm SH}$'s in fcc Pt [Fig.~S6(b)] and the simple cubic $sp$ model [Fig.~3(b)] both exhibit monotonic increase with the orbital texture strength.
Hence, despite its simplicity, the simple cubic $sp$ model reproduces important features of SHE and OHE in Pt and may serve as an illustrative model that captures the key ingredient. 

\subsection{I. \ \ \ \  Boundary accumulation of the orbital and spin moments}

Since the orbital and spin currents are not directly measurable in experiments, we calculate the edge accumulations for the orbital and spin moments in a finite system from the Kubo formula. We considered a film structure with its thickness of 40 atomic layers. In Fig. S7, current-induced orbital and spin moments are shown for Fermi energies $E_\mathrm{F} = -1.30\ \mathrm{eV}$ and $E_\mathrm{F} = 0.00\ \mathrm{eV}$. While the spin and orbital accumulations have the opposite signs for $E_\mathrm{F} = -1.3\ \mathrm{eV}$ [Fig. S7(a)], the signs are the same for $E_\mathrm{F} =  \ 0.00 \ \mathrm{eV}$ [Fig. S7(b)]. This qualitatively agrees with the OHC [Fig.~3(a)] and SHC [Fig.~3(b)] calculated in the bulk.   

\begin{figure}[ht]
	\includegraphics[angle=0, width=0.7\textwidth]{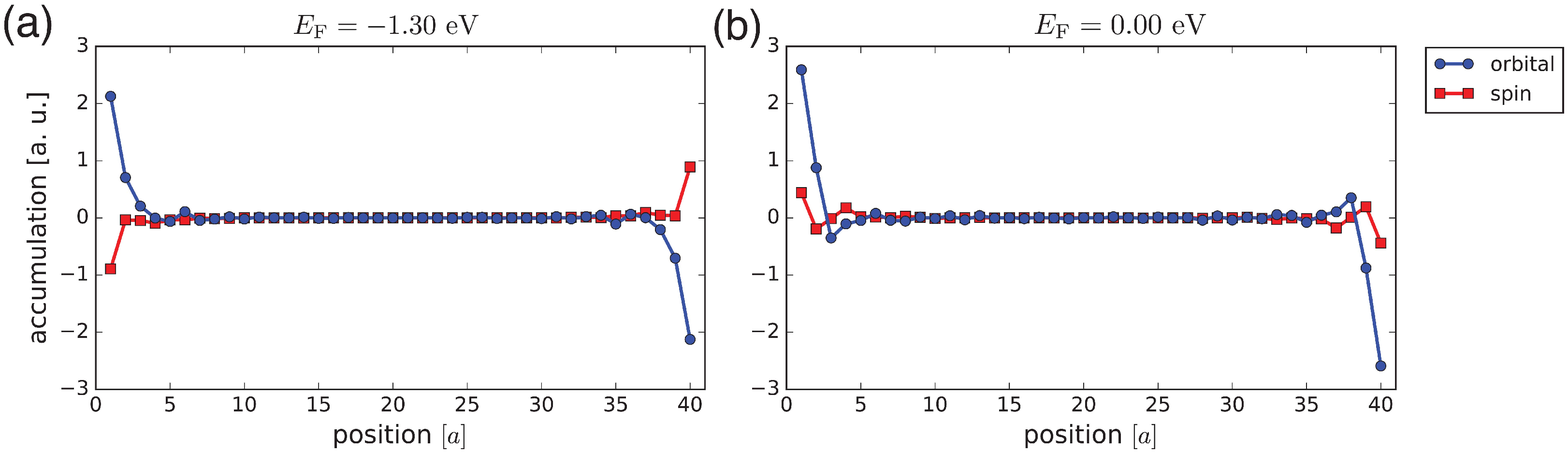}
	\caption{
		Current-induced orbital and spin moments from OHE and SHE in a finite system for (a) $E_\mathrm{F} = -1.30\ \mathrm{eV}$ and (b) $E_\mathrm{F} = 0.00\ \mathrm{eV}$.
	}
\end{figure}

%

\end{document}